\documentclass[apj,twocolumn]{emulateapj}
\usepackage{graphicx, color}
\usepackage{amssymb}
\usepackage[pdftex,colorlinks,hyperfootnotes,citecolor=blue,linkcolor=blue]{hyperref}

\begin{document}
\shorttitle{A candidate $z\sim10$ multiply-imaged galaxy}
\shortauthors{Zitrin et al.}

\slugcomment{Submitted to the Astrophysical Journal Letters}

\title{A Geometrically Supported $\lowercase{z}\sim10$ Candidate Multiply-Imaged by the Hubble Frontier Fields Cluster Abell 2744}

\author{Adi Zitrin\altaffilmark{1,2}, Wei Zheng\altaffilmark{3}, Tom
  Broadhurst\altaffilmark{4,5}, John Moustakas\altaffilmark{6}, Daniel
  Lam\altaffilmark{7}, Xinwen Shu\altaffilmark{8,9}, Xingxing
  Huang\altaffilmark{3,9}, Jose M. Diego\altaffilmark{10}, Holland
  Ford\altaffilmark{3}, Jeremy Lim\altaffilmark{7}, Franz
E. Bauer\altaffilmark{11,12}, Leopoldo Infante\altaffilmark{11}, Daniel D. Kelson\altaffilmark{13}, Alberto Molino\altaffilmark{14}}

%\affil{
\altaffiltext{1}{Cahill Center for Astronomy and Astrophysics,
  California Institute of Technology, MS 249-17, Pasadena, CA 91125,
  USA; adizitrin@gmail.com}
\altaffiltext{2}{Hubble Fellow}
\altaffiltext{3}{Department of Physics and Astronomy, Johns Hopkins
  University, Baltimore, MD 21218, USA}
\altaffiltext{4}{Department of Theoretical Physics, University of
  Basque Country UPV/EHU, Bilbao, Spain}
\altaffiltext{5}{IKERBASQUE, Basque Foundation for Science, Bilbao, Spain}
\altaffiltext{6}{Department of Physics and Astronomy, Siena College,
  Loudonville, NY 12211, USA}
\altaffiltext{7}{Department of Physics, The University of Hong Kong,
  Pokfulam Road, Hong Kong}
\altaffiltext{8}{CEA Saclay, DSM/Irfu/Service d�Astrophysique, Orme
  des Merisiers, 91191 Gif-sur-Yvette Cedex, France}
\altaffiltext{9}{Department of Astronomy, University of Science and
  Technology of China, Hefei, Anhui 230026, China}
\altaffiltext{10}{Instituto de F\'isica de Cantabria, CSIC-Universidad
  de Cantabria, 39005 Santander, Spain}
\altaffiltext{11}{Pontificia Universidad Cat\'olica de Chile, Instituto de
Astrof\'isica, Santiago 22, Chile}
\altaffiltext{12}{Space Science Institute, Boulder, CO 80301, USA}
\altaffiltext{13}{The Observatories of the Carnegie Institution for Sci-
ence, Pasadena, CA 91101, USA}
\altaffiltext{14}{Instituto de Astrof\'isica de Andaluc\'ia - CSIC, Glorieta
de la Astronom\'ia, s/n. E-18008, Granada, Spain}

\begin{abstract}
The deflection angles of lensed sources increase with their distance behind a given lens. We utilize this geometric effect
to corroborate the $z_{phot}\simeq9.8$ photometric redshift estimate of a faint near-IR dropout, triply-imaged by the massive galaxy cluster Abell 2744 in deep Hubble Frontier Fields images. The multiple images of this source follow the same symmetry as other nearby sets of multiple images which bracket the critical curves and have well defined redshifts (up to $z_{spec}\simeq3.6$), but with larger deflection angles, indicating that this source must lie at a higher redshift. Similarly, our different parametric and non-parametric lens models all require this object be at $z\gtrsim4$, with at least 95\% confidence, thoroughly excluding the possibility of lower-redshift interlopers. To study the properties of this source we correct the two brighter images for their magnifications, leading to a SFR of $\sim0.3~M_{\odot}$/yr, a stellar mass of $\sim4\times10^{7}~M_{\odot}$, and an age of $\lesssim220$ Myr (95\% confidence). The intrinsic apparent magnitude is 29.9 AB (F160W), and the rest-frame UV ($\sim1500~\AA$) absolute magnitude is $M_{UV,AB}=-17.6$. This corresponds to $\sim0.1~L^{*}_{z=8}$ ($\sim0.2~L^{*}_{z=10}$, adopting $dM^{*}/dz\sim0.45$), making this candidate one of the least luminous galaxies discovered at $z\sim10$.
\end{abstract}
\vspace{0.2cm}
%\begin{keywords}
%\end{keywords}

%% text by Moustakas
%\def\imtxt#1{{\bf {\textcolor{red}{#1}}}}
%\def\imsout#1{{\bf {\textcolor{red}{\sout{#1}}}}}

\section{Introduction}\label{intro}

The \emph{Hubble} Frontier Fields (HFF; \citealt{Lotz2014AAS_FFreview,
  Coe2014FF}) initiative is a {\em Hubble Space Telescope}
(\emph{HST}) Director's Discretionary Time program in which
4--6 massive galaxy clusters and parallel fields are being observed in the optical and
near-infrared to an unprecedented depth (140 orbits; $m_{\rm AB,
  lim}\sim29$). Coordinated observations with the {\em
  Spitzer Space Telescope} or other ground-base telescopes, and lens-map preparation by several groups, set the stage to advance our understanding of the early Universe. This effort aims at reaching new frontiers of depth into high redshifts, so we may push beyond the current Hubble limit for faint field galaxies to better characterize the properties of the first galaxies and evaluate their role in (re)ionizing the inter-galactic medium.

\begin{figure*}
\vspace{-0.1cm}
\centering
\includegraphics[width=0.9\textwidth]{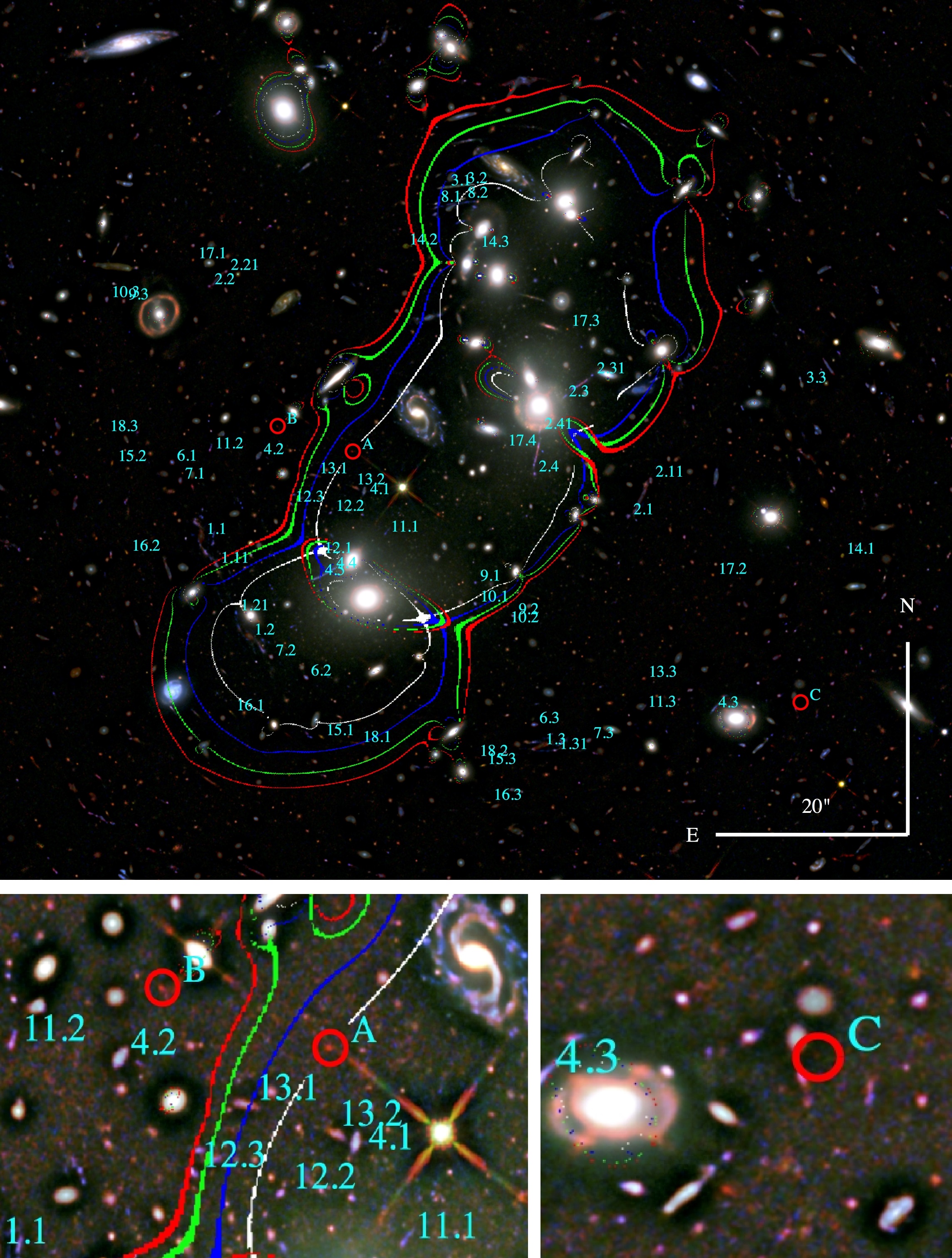}
\caption{Smoothed color mosaic of A2744 (R=F160W+F140W;
  G=F125W+F105W+F814W; B=F606W+F435W) with the expanding critical curves for
  increasing source redshifts (\emph{white}: $z_{s}\simeq1.3$ (system 13); \emph{blue}:
  $z_{s}\simeq2$; \emph{green}: $z_{s}\simeq3.6$ (system 4, \citealt{Richard2014FF}); \emph{red}:
  $z_{s}\sim10$) based on our {\sc ltm} lens
  model overlaid.  The numbered labels indicate the multiple images
  from \citet{Lam2014modelA2744} used as constraints, and
  the \emph{red} circles mark the three images (A, B, and C) of our
  candidate $z\sim10$ dropout galaxy. Our models completely exclude $z\sim2-3$ lower redshifts as a possible solution for this system, as the critical curves should pass midway between the two mirrored images, e.g. A and B here, seen better in the \emph{Bottom left} panel. The \emph{Bottom right} panel similarly shows a zoom-in on our best identification for the least magnified image of this system, image C.}\label{curves0}
\end{figure*}

The HFF will achieve this goal by leveraging the strong gravitational
lensing power of massive galaxy clusters, which deflect, distort,
and---most importantly---magnify distant background galaxies. Thanks to this magnification, background objects that are intrinsically fainter than the observational threshold are magnified in flux and area, at fixed surface brightness, so that the detectability of small, poorly resolved objects is enhanced. Moreover, if the projected surface-mass density of the foreground lens is high enough, multiple images of the same background source often appear \citep[][for reviews]{Bartelmann2010reviewB,Kneib2011review}. For massive cluster lenses many sets of multiple images are typically generated and can be used as constraints for constructing lens models, which can in turn be used to predict the lensing-distance (ratio of the angular-diameter distances of the lens to the source, and the source, $d_{ls}/d_{s}$), and thus constrain the ``geometric redshift'' of other multiply-imaged candidates. The critical curves are bound to expand, and deflection angles increase, with increasing source distance behind a given lens. This ``nesting'' effect \citep[e.g. see related figures in][]{Broadhurst2005a,Zitrin2009_cl0024,Lam2014modelA2744} is especially useful for determining the lensing-distance of high-redshift candidates which happen to be multiply imaged and for which spectroscopy is very challenging \citep[e.g.][]{Vanzella2011Specs7,Bradac2012spec,Finkelstein2013,Schenker2014}.

The magnification by clusters of galaxies has consistently provided galaxies at the highest redshift limits
\citep[e.g.][]{Franx1997,Frye2002,Stark2007,Bradley2008,Bouwens2012highzInCLASH,
  Zheng2012NaturZ}. However, due to the
small source-plane area at high redshifts, the chances of capturing a
multiply imaged high-redshift galaxy are small, with only a few
currently known \citep[e.g.][]{Franx1997, Kneib2004z7,
  Richard2011A383highz, Zitrin2012CLASH0329,Bradley2013highz, Monna2014RXC2248,
  Atek2014A2744, Zheng2014A2744}. The highest-redshift candidate to date was detected to be triply-imaged at $z\sim11$ \citep{Coe2012highz}. While the latter candidate seems secure
in many aspects of its photometric redshift including a scrutinizing
comparison with colors of possible lower-$z$ interlopers, the lens
models could not unambiguously determine its redshift. Similarly, several other $z\sim9-11$ objects are known from deep fields \citep[e.g.][and references therein]{Ellis2013Highz,
  Bouwens2011NaturZ10Gal, Bouwens2014LF, Oeasch2013, oesch14a}, with redshifts
estimated solely on basis of the photometry.

Here, we report a faint, geometrically supported candidate $z\sim10$ galaxy, triply-imaged by the HFF cluster
Abell 2744 (A2744 hereafter). In \S \ref{obs} we summarize the
relevant observations and photometry. In \S \ref{results} we present the photometric
redshifts, lens models, and results, discussed and summarized in \S \ref{Discussion}.
We assume a $\Lambda$CDM cosmology with $\Omega_{\rm
  M}=0.3$, $\Omega_{\Lambda}=0.7$, and $H_{0}=100$ $h$ km
s$^{-1}$Mpc$^{-1}$ with $h=0.7$.

\section{\emph{HST} \& \emph{Spitzer} Observations}\label{obs}

HFF observations of A2744 ($z=0.308$) were obtained between 2013~Oct~25 and 2014~Jul~1 as part of
GO/DD~13495 (P.I., Lotz).  These data consist of 70~orbits with
WFC3/IR in the F105W, F125W, F140W, and F160W near-infrared filters,
and 70~orbits with ACS/WFC in the F435W, F606W, and F814W optical
bandpasses. These observations were supplemented with archival ACS data, $\sim13-16$ ksec in each of these optical filters,
taken between 2009~Oct~27-30 (GO~11689, P.I., Dupke). We also use one orbit imaging in each of the
F105W and F125W bands, and 1.5 orbits in the F160W band, obtained in 2013~Aug and 2014 Jun-Jul (GO~13386; P.I.,
Rodney).

A detailed description of our data reduction and photometry can be found in
\citet{Zheng2014A2744}.  Briefly, both the WFC3/IR and ACS images are
processed using {\tt APLUS} \citep{Zheng2012APLUS}, an automated
pipeline which originally grew out of the {\tt APSIS} package
\citep{Blakeslee2003APSIS}.  We astrometrically align, resample, and
combine all the available imaging in each filter to a common
$0\farcs065$ pixel scale, and create ultra-deep detection images from
the inverse-variance weighted sum of the WFC3/IR and ACS images,
respectively.  The $5\sigma$ limiting magnitude in a $0\farcs4$
diameter aperture in the final WFC3/IR images is approximately
$\sim29$ AB, and $\sim30$ AB in the ACS optical mosaics.

Next, we run {\tt SExtractor} \citep[][]{BertinArnouts1996Sextractor} in dual-image mode
using the WFC3/IR image stack as the detection image.  We require
sources to be detected with a minimum signal-to-noise ratio of 1.5
spanning at least four connected pixels.  We measure colors using
an isophotal aperture defined in the detection image, which balances the
need between depth and photometric precision \citep{Ferguson1995}.
Finally, we identify high-redshift galaxy candidates by looking for a
strong Lyman break using the color cuts given in
\citet{Zheng2014A2744}, supplemented by careful visual inspection. For sources of interest lying near cluster members, such as JD1B and JD1C here (see below), we first run {\tt GALFIT} \citep{PengGalFit2010} to remove the nearby members, before running {\tt SExtractor}. Similarly, for JD1A, a nearby star was removed prior to the photometry (see \S \ref{results}).

In addition to the \emph{HST} observations, we also utilize
\emph{Spitzer}/IRAC imaging of A2744 obtained as
part of Program~90257 (P.I., Soifer) between 2013~Sep and 2014~Feb,
supplemented with archival imaging from 2004 (Program~84; P.I.,
Rieke).  We process the IRAC Basic Calibrated Data (cBCD) images using
standard methods implemented in {\tt MOPEX} \citep{Makovoz2005}, and
create a final mosaic in each channel with a pixel scale of
$0\farcs6$.  The total exposure time of the final mosaics is
$\sim340$~ksec, achieving a $1\sigma$ limiting magnitude of $27.3$ in channel 1 (IRAC1,
$3.6\micron$) and $27.1$ in channel 2 (IRAC2, $4.5\micron$). More details on the IRAC photometry will be given in Huang et al. (in preparation).

\begin{figure}
\centering
\includegraphics[width=90mm]{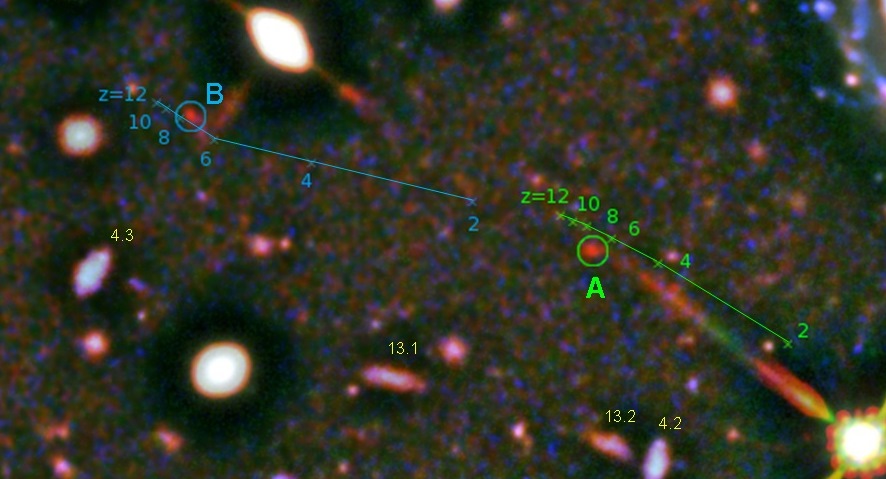}\\
\vspace{0.1cm}
\includegraphics[width=90mm]{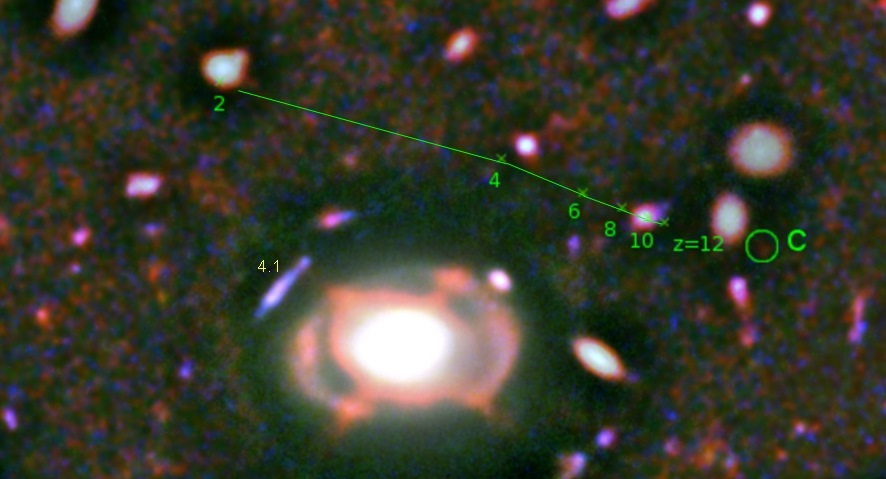}
\caption{\emph{Top:} Loci of predicted positions for images A and B using the \citet{Lam2014modelA2744} model. Images A and B lie close to two other pairs or multiply imaged galaxies at lower redshifts, systems 4 and 13, which also bracket the tangential critical curve (Fig. \ref{curves0}). The \emph{blue} track corresponds to the predicted image position of B using the observed location of image A, and the \emph{green} track is the opposite case. The predictions are shown over a wide redshift range $2<z<12$. High redshift is clearly preferred, explicitly $z>6$, but notice the predicted positions converge at high redshift because of the saturation of the lensing-distance relation (so that a range of high-redshift solutions is allowed). Low redshifts, however, are very clearly excluded. \emph{Bottom:} similar prediction pattern for image C again showing the high-$z$ preference.}
\label{daniel}
\end{figure}

\section{Discovery of The $\lowercase{z}\sim10$ Candidate}\label{results}

We initially identified our high-redshift galaxy candidate as a $J$-band dropout near the center of A2744 (hereafter JD1A).  A preliminary
estimate of JD1A's photometric redshift (see below) suggested it was
most likely at $z\sim10$, although there was a non-negligible
probability of it being a lower-redshift ($z\sim2-3$) interloper.

In order to assess these two possibilities, and motivated by the
vicinity to the critical curves, we
use an updated version of our publicly available {\em light
  traces mass} (LTM) gravitational lensing model of
A2744.\footnote{For details regarding the HFF lens models: \url{ http://archive.stsci.edu/prepds/frontier/lensmodels}} The {\sc ltm}
model assumes that both the baryonic and dark matter mass
distributions can be traced by the cluster's light distribution
\citep{Broadhurst2005a}, where the latter mass component is a smoothed version of the former. This method has been most successful at uncovering large numbers of multiply imaged galaxies in many galaxy clusters \citep[e.g.,][]{Zitrin2009_cl0024, Zitrin2013M0416}, including A2744 \citep{Merten2009}.  Compared to the
publicly accessible {\sc ltm} model of A2744, our new {\sc ltm} model
uses the updated catalog of multiple images
from \citet[][and references therein]{Lam2014modelA2744}, spanning the redshift range $z\sim1-7$
\citep{Atek2014A2744, Zheng2014A2744}.  In Figure~\ref{curves0} we show a color mosaic of the central region of
A2744 together with the critical curves at several different source
redshifts based on our updated {\sc ltm} model, and the
multiple-image constraints from \citet{Lam2014modelA2744}.

Using our {\sc ltm} model, we delens JD1A to the source-plane
and back, considering both the low- and high-redshift hypotheses.  A
source redshift of $z\sim2$ predicts a counter-image $\sim3\arcsec$ northeast of the position of JD1A, and a second counter-image
approximately $20\arcsec$ west of the southern core of the cluster;
however, no other objects are located at either of these predicted
positions within $\sim1-2\arcsec$ (the image reproduction precision \emph{rms} of our
model is $\sim1\farcs3$).  A source redshift of $z\sim10$, on the other hand, yields the
same positional symmetry of multiple images as in the $z\sim2$ case,
but---as expected---with larger deflection angles.  Remarkably, we
find faint $J$-band dropout galaxies near both predicted counter-image
locations (Figure~\ref{curves0}): image B (hereafter JD1B), and, on
the other side of the cluster, a significantly fainter image~C
(hereafter, JD1C), although note that this identification is tentative due to this object's faintness.

We use several independent lensing models to verify the
positions of the predicted multiple images.  First, we construct a second model with the updated \citet{Zitrin2009_cl0024} pipeline, which adopts the LTM assumption only for the galaxies, yet follows an analytical form for the dark matter, namely (projected) elliptical {\sc nfw} distributions
\citep[eNFW;][]{Navarro1996} for the main mass clumps \citep[see][]{Zitrin2013M0416}. This model
(hereafter, {\sc ``nfw''}) is basically identical to the {\sc zitrin nfw}
model released as part of the HFF, but has been updated using the
\citet{Lam2014modelA2744} multiple-image catalog.  Finally, we also check our results against
the lensing model of A2744 supplied by the {\tt CATS} team
\citep[e.g.][]{Richard2014FF}, constructed using the parametric
\texttt{Lenstool} algorithm \citep{Jullo2007Lenstool}, and against the free-form lensing model published by
\citet{Lam2014modelA2744}, which combines both parametric and
non-parametric techniques.

We find that all four lensing models yield consistent results
regarding the predicted multiple image positions of JD1.
Quantitatively, our {\sc ltm} model yields $z\gtrsim4$ for
our candidate, while our {\sc ``nfw''} model requires $z\gtrsim8$, both
with $95\%$ confidence based on more than ten thousand Monte-Carlo
Markov Chain (MCMC) steps. The \citet{Lam2014modelA2744} and \texttt{Lenstool} models, respectively, yield similar results. To illustrate this result, in
Figure~\ref{daniel} we use the \citet{Lam2014modelA2744} model
to plot the predicted positions of images A, B, and C as a function of
source redshift in the range $z=2-12$.  This analysis shows that a
high-redshift solution for our candidate is clearly favored over the
lower-redshift ($z\sim2-3$) alternative.

\begin{figure*}
\centering
\includegraphics[width=1.0\textwidth]{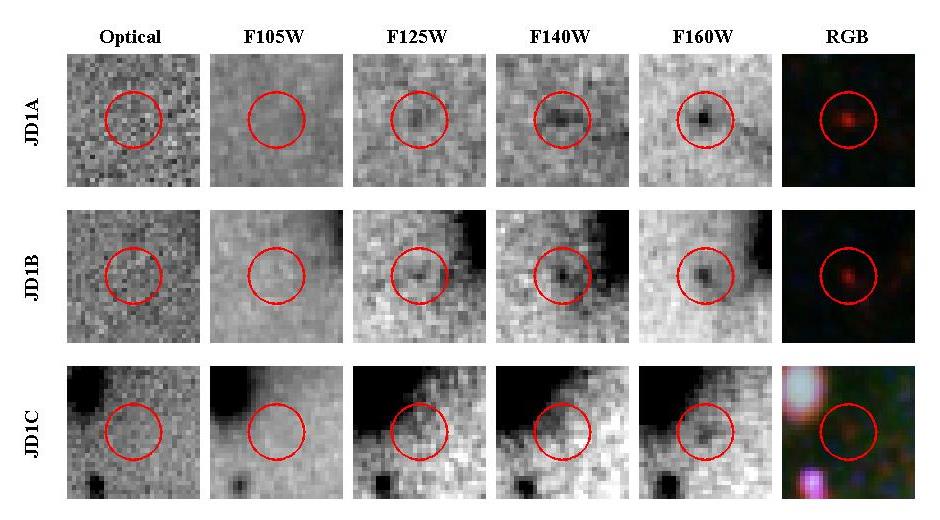}
\caption{Image cutouts of the three multiple images of our $z\sim10$ candidate,
  showing the vanishing flux blueward of the $J_{F125W}$ band. \label{stamps}}
\end{figure*}

Our {\sc ltm} model implies magnifications of
$10.01_{-0.86}^{+1.1}$, $11.25_{-2.5}^{+4.8}$, and
$3.57_{-0.03}^{+0.33}$ ($95\%$ confidence intervals) for JD1A, JD1B,
and JD1C, respectively (Table~\ref{ResultsTable}).
These values are broadly consistent with the magnifications predicted by our updated
{\sc ``nfw''} model, and with the magnifications inferred using the
\texttt{Lenstool} and \citet{Lam2014modelA2744} lensing
models, although these can reach up to $\sim2$ times higher magnifications for images A and B, and $\sim50\%$ lower magnification for image C, and the reader should refer to these as the typical systematic uncertainties here. For calculating the source properties, we shall use the magnifications from our {\sc ltm} model, which renders our calculation conservative in the sense that higher magnifications for A and B yield an even smaller and fainter source than inferred in \S \ref{Discussion}.

In Figure~\ref{stamps} we show $1\farcs74\times1\farcs74$ postage
stamp cutouts of JD1A, JD1B, and JD1C in the observed-frame optical and near-infrared.  JD1A and JD1B are clearly detected in
the three reddest bands (F125W, F140W, and F160W), but vanish in the
F105W bandpass and blueward, suggesting we are observing the
Lyman-break at $z\sim9-10$.  JD1C is significantly fainter, but also
appears to have a Lyman-break at the same observed-frame wavelength.

\begin{figure*}
\centering
\includegraphics[width=0.85\textwidth]{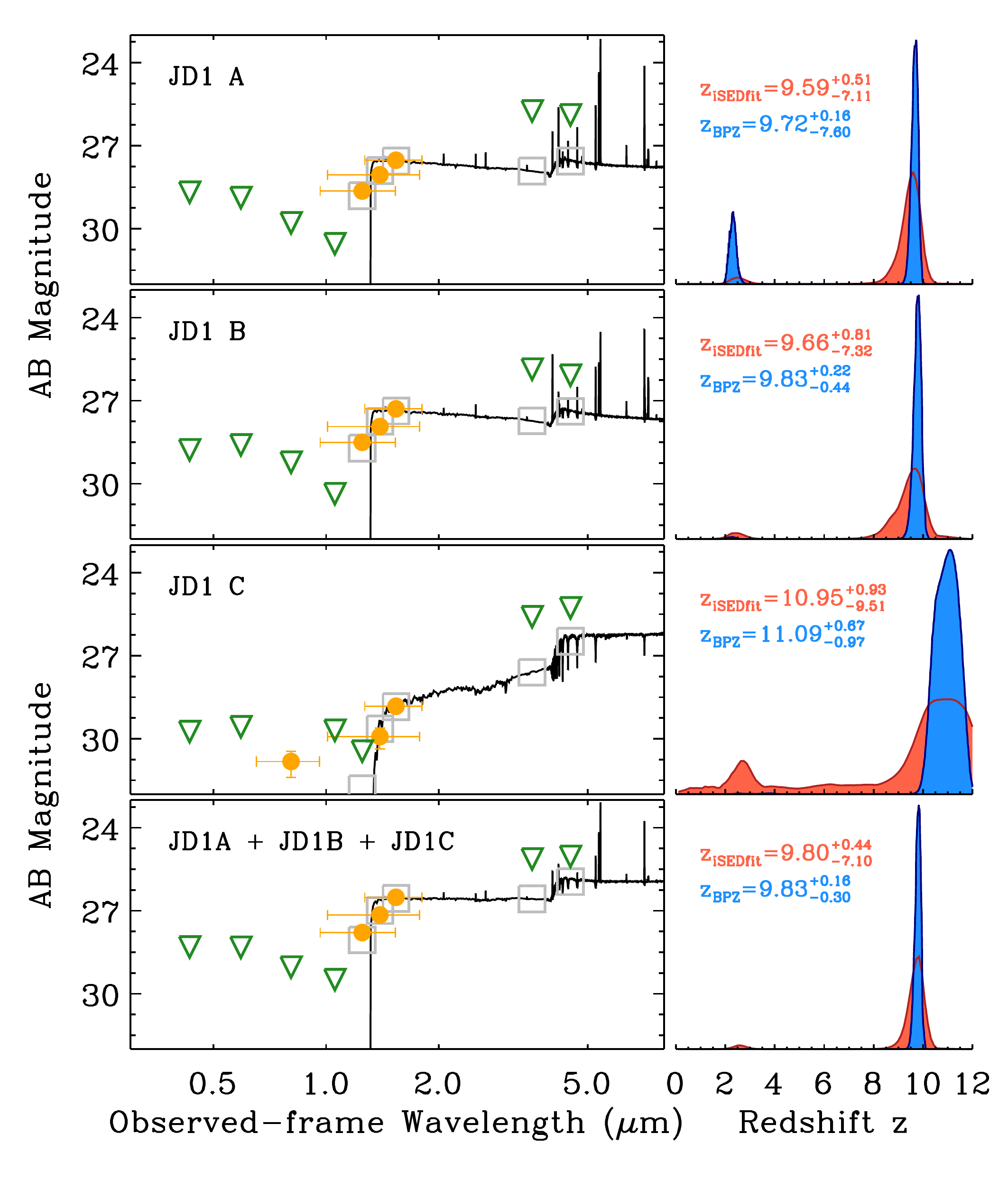}
\caption{\emph{Left:} Observed-frame spectral energy distribution (SED) of JD1A,
  JD1B, JD1C, and of the sum of all three images, JD1A+JD1B+JD1C.  The
  orange points represent statistically significant detections, the
  green triangles are $2\sigma$ upper limits, the black spectrum shows
  the best-fitting (maximum likelihood) model fit from {\tt iSEDfit},
  and the open grey squares represent the best-fitting SED convolved
  with the \emph{HST} filter response curves. \emph{Right:} Photometric redshift probability distribution for each observed SED
  based on both {\tt iSEDfit} (red shading) and {\tt BPZ} (blue
  shading). The legend indicates the redshift at the maximum of
  the marginalized posterior probability distribution, and the 95\% confidence intervals, inferred using
  each code.\label{sedpofz}}
\end{figure*}

We quantify these results in Figure~\ref{sedpofz}, where we show the
spectral energy distributions (SEDs) and photometric redshift
probability distributions of all three sources, as well as the sum
(JD1A + JD1B + JD1C).  We compute photometric redshifts using two
independent codes: {\tt iSEDfit} \citep{Moustakas2013iSED} and {\tt
  BPZ} \citep[Bayesian Photometric Redshifts;][]{Benitez2000,
  Coe2006}.  BPZ relies on a suite of 11 galaxy templates which have
been carefully calibrated using spectroscopic samples over a wide
range of redshift and apparent magnitude.  Like BPZ, {\tt iSEDfit} is
also a Bayesian SED-modeling code, but one which uses stellar
population synthesis models to infer the photometric redshifts and
physical properties of galaxies.  Using {\tt iSEDfit}, we construct a Monte Carlo grid of 20,000 models spanning a wide range of
star-formation history, age, stellar metallicity, dust content, and
nebular emission-line strength (\citealt{Zheng2014A2744} for
details).

Both {\tt BPZ} and {\tt iSEDfit} indicate high-redshift
solutions for both JD1A and JD1B, $z_{\rm phot}\simeq9.6-9.8$, with a secondary small peak at $z_{\rm phot}\sim2.5$. For JD1C,  the preferred redshift is $z_{\rm phot}\simeq11$, with a secondary solution at $z_{\rm phot}\simeq2.5$, although {\tt iSEDfit} finds a
flatter redshift probability distribution allowing less-likely solutions throughout the full redshift range.  Given the
faintness of JD1C compared to the other two images, and with the
supporting evidence from our lensing analysis, we are confident
that JD1 is indeed at $z_{\rm phot}\simeq9.8$.

We conclude this section by describing some additional tests we have
carried out to check the fidelity of our high-redshift candidate.
First, we verify that all three images of our candidate are also present in the
publicly distributed HFF image mosaics, which are independently
processed using the \texttt{MosaicDrizzle} pipeline
\citep{koekemoer11a}.\footnote{\url{http://www.stsci.edu/hst/campaigns/frontier-fields/FF-Data}}
Second, we check the possibility that JD1A may be an artifact of the
nearby stellar diffraction spike (see Figure~\ref{curves0}, although we note that even by eye JD1A is
clearly offset from the diffraction trail).  We select a
comparably bright, isolated star elsewhere in the F160W mosaic and use
its cutout to subtract (after centering and rescaling) the star near
JD1A.  Because the diffraction spikes in the mosaic are all aligned,
this procedure effectively subtracts the offending star and leaves
JD1A unaffected, indicating that it is not an artifact (note, the photometry for JD1A was performed on these star-subtracted images).
As an additional check, we also inspect the archival WFC3/IR imaging
of A2744 from GO~13386 (P.I., Rodney), which is rotated by
approximately $9^{\circ}$ relative to the HFF mosaics, and find that
both JD1A and JD1B are present (although only within the noise level due to
the shallowness of this imaging), again suggesting these are not artifacts related to the spikes.  Finally, we verify that neither
JD1A nor JD1B are moving, foreground objects by creating custom
mosaics from the first and second half of the individual F160W
exposures obtained as part of the HFF observations.  JD1A
and JD1B are both clearly detected in both mosaics.  Furthermore,
subtracting the two mosaics causes both sources to disappear, again
indicating that these are bona fide extragalactic sources.

\section{Discussion \& Conclusions}\label{Discussion}

We report the discovery of a $z\sim10$ Lyman-break galaxy
multiply imaged by the massive galaxy cluster A2744, which has been
observed to an unprecedented depth with \emph{HST} as part of the HFF
campaign.  This candidate adds to just several other
galaxies reported to be at $z\sim9-11$ (see \S 1), and
therefore provides important insight into galaxy formation at the
earliest epochs. Despite lack of spectroscopy for such high-redshift objects, with a variety of well-constrained lens models we are
able to geometrically confirm that this object must lie at high
redshift.

To constrain the physical properties of our candidate, we fix the redshift at the most probable redshift, $z_{\rm phot}=9.8$,
and use {\tt iSEDfit} to construct a large suite of model SEDs. After accounting for the individual magnifications of each image (see Table 1), we find that JD1 has a stellar mass of $\sim4\times10^{7}$~M$_{\odot}$ and is forming stars at approximately $0.3$~M$_{\odot}$~yr$^{-1}$, implying a doubling time\footnote{The time it would take for the galaxy to double its stellar mass, assuming a 25\% gas loss factor appropriate for a $\sim200$~Myr stellar population \citep{behroozi13}.} of $\sim500$~Myr, comparable to the age of the Universe at $z=9.8$.  Using the two brightest sources (JD1A and JD1B), we are also able to constrain the SFR-weighted age to $<220$~Myr (95\% confidence), implying a formation redshift of $z_{f}<15$.

To examine the intrinsic size of the galaxy we focus on JD1A. We measure an approximate
half-light radius of $\sim0.1\arcsec$ in the image plane, corresponding to a
delensed half-light radius of $\lesssim0.03\arcsec$ ($\lesssim0.13$
kpc). This source size is several times smaller than expected
following recent $z\sim9-10$ candidates uncovered in deep
fields \citep[e.g.,][]{Ono2013sizes_highz,oesch14a,Holwerda2014highz}, or, following the
observed size evolution extrapolated from lower-redshifts (\citealt{Coe2012highz}, and references therein). However, the
source size we find matches very well the size of the lensed
$z\simeq10.7$ candidate galaxy published in Coe et al., who showed that although being smaller than expected by a factor
of a few, the typical factor $\sim2$ scatter in sizes found in somewhat
lower-redshift galaxies, alleviates the
discrepancy. It is interesting that both these
highest-redshift multiply-imaged candidates to date, exhibit smaller
sizes than high-$z$ field objects.

Finally, the magnification by our lens models indicate that the intrinsic apparent magnitude is 29.9 AB (F160W), and the rest-frame UV ($\sim1500~\AA$) absolute magnitude is $M_{UV,AB}=-17.6$, corresponding to $\sim0.1~L^{*}_{z=8}$, or $\sim0.2~L^{*}_{z=10}$ (extrapolated with $dM^{*}/dz\sim0.45$). This makes this galaxy one of the least luminous $z\sim10$ candidates ever discovered, supplying a first taste of the upcoming achievements of the HFF observational effort -- reaching deeper into the faint end of the high-redshift luminosity function.

\section*{acknowledgments}
We kindly thank the anonymous reviewer of this work for useful comments. Useful discussion with Richard Ellis, Rychard Bouwens, and Larry Bradley, are much appreciated. This work is based on observations made with the NASA/ESA {\it Hubble Space Telescope} and was supported by
award AR-13079 from the Space Telescope Science Institute (STScI),
which is operated by the Association of Universities for Research in
Astronomy, Inc. under NASA contract NAS~5-26555. It is also based on observations from the {\it
  Spitzer Space Telescope} which is operated by the Jet Propulsion
Laboratory, California Institute of Technology under a contract with
NASA.  This work utilizes gravitational lensing models produced by
P.I.s Zitrin \& Merten, and P.I. Ebeling, funded as part of the HST
Frontier Fields program conducted by STScI. Support for this
work was provided by NASA through Hubble Fellowship grant
\#HST-HF-51334.01-A awarded by STScI.

\bibliographystyle{apj}
\bibliography{outDan3Agnese}

\clearpage

\setlength{\tabcolsep}{0.02in}
\begin{deluxetable}{ccccccccccccc}
\tablecaption{Photometry, Redshifts, and Magnifications of our
  $z\sim10$ Candidate\label{ResultsTable}}
\tabletypesize{\scriptsize}
\tablewidth{0pt}
\tablehead{
\colhead{} &
\colhead{R.A.} &
\colhead{Dec.} &
\colhead{F160W} &
\colhead{F140W} &
\colhead{F125W} &
\colhead{F105W} &
\colhead{F814W} &
\colhead{$3.6\micron$\tablenotemark{a}} &
\colhead{$4.5\micron$\tablenotemark{a}} &
\colhead{} &
\colhead{{\tt iSEDfit}} &
\colhead{{\tt BPZ}} \\
\colhead{Name} &
\colhead{(J2000, deg)} &
\colhead{(J2000, deg)} &
\colhead{(nJy)} &
\colhead{(nJy)} &
\colhead{(nJy)} &
\colhead{(nJy)} &
\colhead{(nJy)} &
\colhead{(nJy)} &
\colhead{(nJy)} &
\colhead{$\mu_{LTM}$\tablenotemark{b}} &
\colhead{Redshift\tablenotemark{c}} &
\colhead{Redshift\tablenotemark{c}}
}
\startdata
JD1A & $3.59251$ & $-30.40149$ & $35.4\pm1.4$ & $21.8\pm1.7$ &
$12.7\pm1.4$ & $1.2\pm1.1$ & $4.3\pm2.2$ & $<91$ & $<80$ &
$10.01_{-0.86}^{+1.1}$ & $9.59^{+0.51}_{-7.11}$ & $9.72^{+0.16}_{-7.60}$
\\
JD1B & $3.59502$ & $-30.40075$ & $43.8\pm2.9$ & $24.3\pm2.5$ &
$14.4\pm2.0$ & $2.0\pm1.3$ & $2.1\pm3.7$ & $<82$ & $<67$ &
$11.25_{-2.5}^{+4.8}$ & $9.66^{+0.81}_{-7.32}$ &
$9.83^{+0.22}_{-0.44}$ \\
JD1C & $3.57753$ & $-30.40871$ & $10.7\pm0.7$ & $3.9\pm1.3$ &
$0.7\pm1.2$ & $-2.0\pm2.4$ & $1.7\pm0.7$ & $<105$ & $<141$ &
$3.57_{-0.03}^{+0.33}$ & $10.95^{+0.93}_{-9.51}$ & $11.09^{+0.68}_{-0.97}$
\\
Total\tablenotemark{d} & \nodata & \nodata & $89.9\pm3.3$ & $50.0\pm3.3$ &
$27.8\pm3.1$ & $1.2\pm2.9$ & $8.1\pm4.4$ & $<161$ & $<175$ & \nodata
& $9.80^{+0.44}_{-7.10}$ & $9.83^{+0.16}_{-0.30}$
\enddata
\tablenotetext{a}{Local limiting fluxes ($\sim1\sigma$ upper limits) estimated by running {\tt GALFIT} to model the IRAC images in a  $\sim13\arcsec \times13\arcsec$ region around each position to subtract all nearby sources, and measuring the \emph{rms} fluctuation of the residual images.}
\tablenotetext{b}{Magnifications and $95\%$ confidence intervals from our {\sc ltm} gravitational lensing model.}
\tablenotetext{c}{Uncertainties on the photometric redshifts are
  $95\%$ confidence intervals.}
\tablenotetext{d}{Summed flux of all three sources (JD1A+JD1B+JD1C)
  with the uncertainties added in quadrature.}
\end{deluxetable}

\end{document}